# Web Search Result Clustering based on Cuckoo Search and Consensus Clustering

Mansaf Alam and Kishwar Sadaf
Dept. of Computer science,
Jamia Millia Islamia, New Delhi, India

**Abstract**

Clustering of web search result document has emerged as a promising tool for improving retrieval performance of an Information Retrieval (IR) system. Search results often plagued by problems like synonymy, polysemy, high volume etc. Clustering other than resolving these problems also provides the user the easiness to locate his/her desired information. In this paper, a method, called WSRDC-CSCC, is introduced to cluster web search result using cuckoo search meta-heuristic method and Consensus clustering. Cuckoo search provides a solid foundation for consensus clustering. As a local clustering function, k-means technique is used. The final number of cluster is not depended on this k. Consensus clustering finds the natural grouping of the objects. The proposed algorithm is compared to another clustering method which is based on cuckoo search and Bayesian Information Criterion. The experimental results show that proposed algorithm finds the actual number of clusters with great value of precision, recall and F-measure as compared to the other method.

# 1. Introduction

Today, Information Retrieval (IR) systems like search engines play an important role in lives of people who seek information from the web. The lack of standard and uniformity makes retrieval of information challenging. Furthermore the queries placed by naïve users also make retrieval even more complicated job. To cover all those issues, a general search engine delivers millions of documents in an answer to queries that may be specific or ambiguous. This voluminous set of search results cannot be grasped by a user. And more importantly, the desired information may not be there in the first display of the search engine window. Moreover the first few results are dominated by frequently search pages or documents. For instance, for the query "puma", a conventional search engine returns result whose leading pages are related to "puma brand" irrespective of the user's requirement. "Puma" query may imply multiple things like animal or web server. Hence a user has to traverse the search result, which Höchstötter et al [1] noted that users seldom traverse all the pages of search result. Their searching is limited to only first few pages of the search engine. Therefore a method is needed to organize this gigantic search result. A way to organize this result set is to cluster them into thematic groups. Clustering is a technique of grouping objects where objects of one group are similar to each other and dissimilar to objects of other groups. Method of clustering of search result was proposed way back in the year 1992 by Cutting et al [2] in their Scatter/Gather system. This system is held as the conceptual father of all search result clustering systems. There are many clustering engines commercially available like Carrot2, Kartoo, DuckDuckGo etc. Details about various search result clustering methods and clustering engines can be found in [3].

Clustering is an unsupervised machine learning technique. The unsupervised feature makes it more fitting for clustering search result as it is not possible to determine as to how many categories are there in search result. Clustering of web search involves four basic steps [4]: a) search result acquisition, b) result preprocessing, c) cluster formation and d) labeling of clusters. Some clustering engines acquire search results from one or more search engines and then merge them into one unified result set. There are some clustering engines too which acquire search result by actually employing their own searching method. In preprocessing, each and every document of page of the search result is transformed into streams of words or phrases or sentences depending upon the attributes of the clustering method. Other tasks performed during pre-processing are stop word removal, stemming, filtering etc. cluster formation and labeling may go hand in hand.

The objective of search result clustering is to provide users easiness in locating their information need rather than improving the ranking of documents in the result. A most common approach to cluster web search result is to cluster document snippets returned with each URL in the result. Suffix Tree Clustering (STC) is one of the most popular snippet clustering method proposed by Zamir et al [5]. The similarity between documents are measured using common phrases instead of single word. This technique inspired many other notable works [6, 7, 8]. To cluster search result, Mecca et al [9] perform their method on the whole document instead of small snippets. In addition to text-based methods, clustering can be performed using the hyperlink structure of the documents of the result [10, 11, and 12]. There are many studies dedicated to the problem of search result clustering [13, 14, 15, and 16]. The proposed approach is based on the textual contents of the result. It is based on meta-heuristic cuckoo search, k-means and consensus clustering. Cuckoo search with k-means provide solutions that are locally and globally balanced. The solutions which are produced by cuckoo search are used as input to the consensus clustering method. Consensus clustering is a method that finds a consensus among multiple clusterings performed on the datasets. The advantage of proposed algorithm is that it can find the actual number of clusters present in the search result. Meta-heuristic methods like Genetic Algorithm (GA), Particle Swarm Optimization (PSO) have been utilized in document clustering. Kamel et al in [17], propose a document clustering approach which is based on PSO and k-means. Wei et al in [18], proposed a document clustering approach using genetic algorithm and Latent Semantic Indexing (LSI). In [19], authors show that cuckoo search outperforms GA and PSO in clustering their benchmark datasets. As per our knowledge, the proposed method which is based on cuckoo search, k-means and consensus clustering is the first study in context of search result clustering.

The rest of the paper is organized as follows. Section 2 reports some state of art in the area of web search result clustering. Proposed method is described in detail in section 3. Experimentation and results are discussed in section 4. Page and documents are used interchangeably throughout the paper.

## 2. Related Work

Search result clustering can be categorised into different types from different perspectives. Traditional clustering can be broadly categorised into hierarchical and flat clustering. [20, 21]. Hierarchical clustering are subdivided into two classes: agglomerative and divisive. Agglomerative clustering starts by assigning each object to different singleton cluster. In each iteration, these clusters are merged based on some similarity metric until a stopping criteria is reached. On the contrary, in divisive approach, initially all objects are assigned to one cluster. In each iteration, this cluster is divided into different clusters. This process continues until a stopping criteria is met. In flat or partitional clustering, objects are grouped in a single go. K-means and its variant k-mode, spherical k-means are some examples of flat clustering. Partitional and hierarchical clustering using textual contents of the objects are most common algorithms for search result clustering [22]. Although partitional clustering methods are faster than the hierarchical methods, but initial bad seed choice can severely degrade the performance.

The documents of web search result have distinct features other than their textual contents like hyperlinks. Clustering can be classified as graph-based, rank-based etc. [23]. In [24], a method to cluster search result, Lingo, is proposed which is based on Latent Semantic

Indexing (LSI) using snippet phrases. This method has been incorporated in Carrot2 clustering engine. Another famous mathematical technique Non- negative Matrix Factorization (NMF), have been utilized in clustering documents [25, 26]. NMF is an improvement over LSI. It preserves the data non-negativity besides providing low rank approximation of large term-document matrix. In [27], a search result clustering algorithm is proposed which is based on NMF, where NMF is employed in decomposing the term-snippet matrix to produce labels. The snippets are then matched with labels to create clusters. The work for performing the heuristic search and LSI proposed [40] that method handles the heuristic search on query result graph that is to prune undesired edges to build cluster as well as carries out Latent Semantic Indexing boundary clusters to prepare them polished, meaningful, and applicable to query.

## 2.1 Cuckoo Search

Cuckoo Search is a meta-heuristic search method proposed by Xin-She et al in [28]. It mimics the parasitic breeding behaviour of cuckoos. Cuckoos do not breed their eggs. A cuckoo relies on other birds to host its egg. To do so, a cuckoo first selects a random nest and lays its eggs there. The host bird after finding an alien egg may destroy it or abandon the nest. In order to avoid the detection, cuckoos emulate the size, colour and shape of the eggs of hosting bird. The search for a nest by cuckoos follows the Levy flight distribution.

The cuckoo search algorithm characterizes all these behaviours. The breeding behaviour of cuckoos can be summed up in three rules [28]: (i) each cuckoo lays one egg at a time and places it in a randomly selected nest; (ii) nests with high quality eggs would be carried to nest level production; (iii) the number of nests is fixed and the probability of discovery of cuckoo egg by the host bird is $p_a[0, 1]$. The host bird either destroys the egg or abandon the nest and build a new nest. The basic cuckoo search algorithm is based on these three rules.

1. Generate n nests
2. Repeat till stopping criteria is met
    a. Randomly get a cuckoo to a nest using Levy flight
    b. Evaluate its quality ($F_i$)
    c. Randomly select a nest
    d. Evaluate its quality ($F_j$)
    e. If $F_i > F_j$ then replace $j$ with new solution
    f. A fraction $p_a$ of worst nets are replaced by new nests
    g. Calculate the quality of nests and keep the best nests
    h. Rank the nest.
3. Process the result

We adopted the cuckoo search method to create multiple solution sets and then find a consensus among the nests using consensus clustering.

## 2.2 K-means Clustering

K-means is the most famous flat clustering technique that has been used extensively in the field of IR. Although this method is not specific for clustering documents, yet it has been widely practiced due to its linear complexity. It groups data objects into k clusters. To find k distinct clusters, it uses Sum of Squared Error criteria method.

$$E = \frac{1}{n}\sum_{i=1}^{k}\|d - c_i\|^2 \tag{1}$$

Where $c_i$ is the closest centroid to object $d$ and $n$ is total number of objects. The biggest drawback of this method is the user selection of value k. A bad choice of this k would lead to unexpected and inaccurate grouped clusters which is highly undesirable. Many studies are available in literature that propose methods to estimate the value of k [29, 30]. In [31], authors propose a method for clustering documents based on harmonic search and k-means. They show that their algorithm achieves best solutions even with bad selection of value k. After spotting k centres, objects are assigned to centroids with which they have smallest Euclidean distance. This distance depicts the measure of similarity. In our algorithm, we employ the Cosine similarity method. To measure the similarity between two objects, cosine method calculate the orientations of the objects. For instance, Euclidean method of distance may place two documents having word "cow" 50 and 500 times respectively, far away from each other irrespective of their similar topic. But cosine method finds the orientation of the documents which points to similar direction.

## 2.3 Consensus Clustering

As the name suggests, consensus clustering is finding a clustering solution in a set of clusterings that is in the agreement with them. Consensus clustering, in literature, is also known as clustering aggregation, clustering ensembles, and clustering combinations. It can be considered as a meta-clustering technique where multiple clusterings whether same or different, are combined to give an optimal clustering solution. This topic has attracted many studies in variety of areas [32, 33, 34, 35, 36, 37]. Given a set of *m* clusterings $Z_1$, $Z_2$,.., $Z_m$, consensus clustering tries to find a clustering Z that is in the least disagreement with the *m* clustering. Consensus clustering not only finds a consensus among multiple clusterings but also reveals the organizational differences presented in the given clusterings.

### 2.3.1 Consensus Clustering Formulation

Consider a set of data objects $D = \{d_1, d_2, ..., d_n\}$. Let $P = \{p_1, p_2, ..., p_m\}$ represents $m$ clusterings results performed over $D$. Consensus clustering tries to find a clustering $p^*$ that is in agreement with the $m$ clusterings. Many methods have been proposed to cluster results of multiple clusterings on the same data set.

In [38], authors explain two scenarios out of many when multiple clusterings are performed on the same data set. The first scenario is clustering large categorical data. For example, a database that is defined over many categories. Applying clustering multiple times on various combinations of categories yields different results. Consensus clustering combines all these clustering solutions to achieve a clustering which is in least disagreement with the input clusterings. The second situation arises when a clustering system returns different results in each run on the same dataset. Non-deterministic clustering method like k-means yield different result on the same input when run several times. Consensus clustering derives a consensus among the results produced during multiple runs of k-means by producing a strong clustering solution. In this light, we employ consensus clustering on solution nests, each representing results of k-means clustering.

Given a set of clusterings $P = \{p_1, p_2, ..., p_m\}$ defined over dataset $D = \{d_1, d_2, ..., d_n\}$, to find consensus, the distance between these clusterings can be calculated. Let $X$ denotes the distance between two clusterings. The distance between two clusterings is calculated by counting pairwise similarity or dissimilarity of objects of $D$ as:

$$X_{u,v}(p_i, p_j) = \begin{cases} 1 & \text{if } p_i(u) = p_i(v) \text{ and } p_j(u) \neq p_j(v) \\ & \text{or } p_i(u) \neq p_i(v) \text{ and } p_j(u) = p_j(v) \\ 0 & \text{Otherwise} \end{cases} \quad (2)$$

$$S(p_i, p_j) = \sum X_{u,v}(p_i, p_j) \quad (3)$$

Where *u* and *v* are consecutive objects in *D*. When consensus clustering is used to minimize *S*, it is also known as Median Partition problem. Consensus clustering problem is known to be NP-complete. Median partition problem finds a clustering or partition p that has the minimum distance from all the given input clusterings. Since the distance between clusterings can be defined in several ways, finding an optimal clustering is NP-Complete problem. Several heuristic methods have been proposed to reduce this problem into computable form [34, 38]. Local search is one such method. It is a meta-heuristic search method which finds optimal solution by moving from solution to solution in a solution space.

## 3. Proposed Method: WSRDC-CSCC

We propose an algorithm called Web Search Result Document Clustering based on Cuckoo Search and Consensus Clustering (WSRDC-CSCC) for clustering web search result which is based on cuckoo search, a powerful meta-heuristic search method and consensus clustering. K-means is used as a local optimizer. Studies have shown that cuckoo search performs better than other meta-heuristic methods like Particle Swarm Optimization or Genetic Algorithm [39, 19]. Cuckoo search provides a solid foundation for employing consensus clustering. Instead of calculating quality of each nest, we exploit consensus clustering to find an optimal solution. Our work is particularly inspired by the method of cuckoo search proposed by [4]. The authors adopted cuckoo search in clustering the web search result. In their method, the Levy Flight behaviour of cuckoos is transformed into creation of new nest and split-merge operations of clusters. The probability PA of abandoning a nest is given by (real values between 0.1 and 0.2). We utilize these functions in our study.

### 3.1 Problem Formulation

Given a set of search result $D = \{d_1, d_2, \dots, d_n\}$, we apply Cuckoo search on D using k-means for creating initial clusters. The number k is selected as chosen by [4]:

$$k = trunc\ |\sqrt{n} + 1| \tag{4}$$

Where n is the number of documents in the result set. The advantage of our approach is that the number of clusters is not depended on this k. Consensus clustering finds the natural groupings irrespective of k. The solution nests produced by cuckoo search method are uses as input to the consensus clustering.

---

Input: $D = \{d_1, d_2, ..., d_n\}, n = |D|$

Output: $C = \{c_1, c_2, ..., c_l\}$

Step 1: Initialize Specific number of Nests containing D. $P = \{p_1, ..., p_m\}\ where\ p_i = \{d_1, ..., d_n\}$

Step 2: Perform k-means on each nest until convergence.

Step 3: Create a new nest in terms of abandonment, split or merge operations of a randomly chosen nest.

Step 4: Perform consensus clustering on solutions given by each and every nests.

Step 5: Perform Local Search to get the final optimized solution.

---

Figure 1: Algorithm for WSRDC-CSCC

The details of above algorithm are as follows:

Step 1: **Initialize the nests.** In this step nests are initialized. In this study, we select 5 nests initially. Each nest contains a copy of D. Each document of D is pre-processed and converted into vector form.

$$D = \{d_1, d_2, .., d_n\} \tag{5}$$

Step 2: **Perform K-means**. In this phase, we apply k-means in each nest for specific number of times to prevent k-means from converging too quickly. The Euclidean distance metric is used to measure the similarity between the document vectors. At the end of this step, k clusters are produced in each nest. Let C is the set of clusters created in each nest:

$$C = \{c_1, c_2, .., c_k\} \quad (6)$$

$$\cup_{i=1}^{k} c_i = \{d_1, \dots, d_n\} \text{ and } c_i \cap c_l = \emptyset \text{ and } i \neq l \quad (7)$$

Step 3: **Creating a new nest**. To simulate the Levy Flight behaviour, abandonment, creation of new nest, split and merge operations are performed in this phase. We randomly select a nest from the pool of nests and copy its contents to create a new nest. In this nest, the most dispersed cluster is split to create new clusters. In merge operation, two most similar clusters are merged.

Step 4: **Consensus Clustering**. Instead of calculating quality of each nest to find the best nest, we apply consensus clustering on all the clusterings represented by nests. Lets P represents a set of m nests or clusterings, $P = \{p_1, \dots, p_m\}$ where $p_i = \{c_1, \dots c_k\}$. Whatever is the value of k, the final number of clusters depends on the consensus clustering. To reach a consensus among P nests, we find the documents on which m nests agree. Lets u and v are two documents of D. We count the times these two are clustered together in P. Given two nests $p_1$ and $p_2$, we count the times when two objects are co-clustered. Lets T be the similarity measure represented by

$$T_{u,v}(p_1, p_2) = \begin{cases} 1 \text{ if } p_1(u) = p_1(v) \text{ or } p_2(u) = p_2(v) \\ 0 \text{ otherwise} \end{cases} \quad (8)$$

To compute overall consensus between p1 and p2, we count the pair-wise similarity between objects of D by counting how many times they are grouped together. If two objects are co-clustered in majority of m clusterings, then they are clustered together. Let $M = n \times m$, a matrix where n is the number of documents and m is the number of clusterings.

$$M = \begin{pmatrix} p_1(d_1) & \cdots & p_m(d_1) \\ \vdots & \ddots & \vdots \\ p_1(d_n) & \cdots & p_m(d_n) \end{pmatrix} \qquad (9)$$

From M, we can create a *m* dimensional vector $v_i$ to corresponding $d_i$.

$$v_i = [p_1(d_i), p_2(d_i), \ldots, p_m(d_i)] \qquad (10)$$

The consensus is reached between u and v when their corresponding vectors satisfy $\alpha(|T_{u,v}|/m > 0.5)$, i.e. when majority of clusterings agree. Let the ensemble clustering of m clusterings be denoted by $\pi = \{c_1, \ldots, c_l\}$. The documents are clustered in the following mode.

---

**Consensus Clustering**

---

**For all pairs** $(d_i, d_j)$

    ***if*** $\alpha\left(d_i(v), d_j(v)\right) = $ ***True***

        ***Create*** $c_a$ ***and*** $c_a \leftarrow \{d_i, d_j\}$

    ***else*** $c_a \leftarrow \{d_i\}$,

    ***Create*** $c_b$ ***and*** $c_b \leftarrow \{d_j\}$

---

## 4. Experimentation and Result

The proposed method of web search result clustering using Cuckoo search and consensus clustering was tested on the datasets available on

http://artemisa.unicauca.edu.co/~ccobos/wdc/wdc.htm. The datasets comprised of web search results from Dmoz in response to 50 different queries. Each query result has on average 129.14 documents. We considered 5 datasets from the above mentioned directory. We set the same parameters values as considered by [4] in their method. We fixed the number of nest population as 5, and number of iteration as 4 in k-means phase. Also instead of Euclidean distance, we employed cosine similarity for assigning objects to centroids in k-means method. The computation of k-means in each nest can be executed in parallel manner. We adopted the approach of split, merge and creation of new nest to simulate the behaviour of Levy flight as proposed by [4].

The result of our proposed method is remarkable as it exactly found the actual number of clusters present in the datasets. Irrespective of value of k, our method correctly identified the real clusters with good precision and recall. Although consensus clustering can be directly applied on the clustering solutions represented by nests, split and merge operations are crucial in providing a sound base for consensus clustering. The creation of new nest to show abandonment of a nest in cuckoo search is crucial to our consensus clustering as the split and merge operations are performed in this new created nest. The split and merge functions increased the quality of consensus clustering. We found that consensus clustering without split and merge function, failed to produce good and actual number of result. Also it is observed, the more number of split, and merge methods are performed on the copy of a randomly selected nest, the more quality clusters got produced in consensus clustering. We considered each and every produced cluster as different search result returned by the search engine in an answer to a query. And the total number of relevant documents, pertaining to each cluster, in the whole result set is considered as the corresponding class. To evaluate the quality of produced clusters, we utilised famous IR evaluation metrics: precision, recall, and F-measure.

## 4.1 Comparison

Our algorithm, WSRDC-CSCC, was compared with a method proposed by Cobos et al [4], which is based on Cuckoo search, k-means and Bayesian Information Criterion (BIC) called WDC-CSK. The advantage of our proposed approach is the identification of actual number of clusters. For comparison purpose, we applied WDC-CSK BIC on our datasets. Authors in [4], first create a pool of nests containing a copy of dataset. Then in each nest, k-means is performed. To show the Levy Flight behaviour, a nest is randomly selected and its solution is copied to create a new nest. The split and merge functions are performed in the newly created nest. Authors show that the new nest created to simulate the behaviour of Levy Flight, provides the best solution. The more iterations of split and merge are performed, the more effective clusters are produced. Although this method produced good quality clusters, but it could not discover the actual number of clusters. Table 1 shows the final number of clusters produced by both methods and the difference to actual number of clusters present in the datasets. Moreover WSRDC-CSCC achieved good precision and recall values as compared to WDC-CSK BIC approach. Figure 2 and Figure 3 present the precision, recall and f-measure graph for both the methods.

| Dataset | Algorithm | Estimated k | Difference to actual k |
|---------|-----------|-------------|------------------------|
| DS1 | WSRDC-CSCC | **4** | **0** |
|  | WDC-CSK BIC | 9 | 5 |
| DS2 | WSRDC-CSCC | **6** | **0** |
|  | WDC-CSK BIC | 8 | 2 |
| DS3 | WSRDC-CSCC | **5** | **0** |
|  | WDC-CSK BIC | 10 | 5 |
| DS4 | WSRDC-CSCC | **4** | **0** |
|  | WDC-CSK BIC | 8 | 4 |
| DS5 | WSRDC-CSCC | **6** | **0** |
|  | WDC-CSK BIC | 8 | 2 |

Table 1: Estimated number of cluster and difference to actual number of clusters

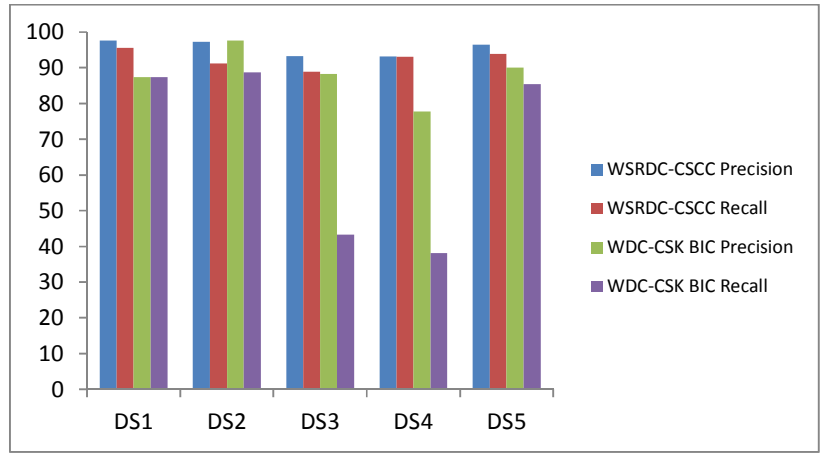

Figure 2: Precision and recall graph for WSRDC-CSCC and WDC-CSK BIC

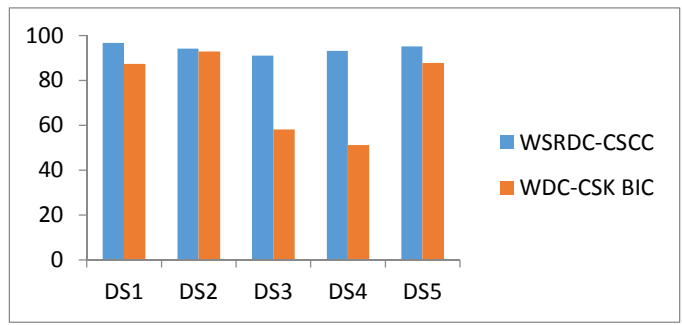

Figure 3: F-measure graph for WSRDC-CSCC and WDC-CSK BIC

## 5. Conclusion

The proposed algorithm for clustering web search result is based on cuckoo search and consensus clustering. K-means has been used as a local improvement function. The experimental result shows that our method achieves good precision, recall and f-measure values. The most remarkable feature of our algorithm is the identification of actual number of clusters present in dataset. The result also depends on how many times the split and merge functions are performed on the clusters of a randomly created nest. We found that creation of new nest creation with split and merge operations improve the quality of consensus clustering. To reach consensus among clusterings represented by nests, a pair-wise similarity method has been applied on these clusterings. If two objects are co-clustered in majority of clusterings, then they are clustered together in the output clustering solution. As a future work, we identified several tasks, for example labeling of search result using some external lexicon and clustering of search result based on ontologies.